# The Physical Web in Smart Cities

Dmitry Namiot
Faculty of Computational Mathematics and Cybernetics
Lomonosov Moscow State University
Moscow, Russia
dnamiot@gmail.com

Manfred Sneps-Sneppe
Ventspils International Radio Astronomy Centre
Ventspils University College
Ventspils, Latvia
manfreds.sneps@gmail.com

*Abstract*— In this paper, we discuss the physical web projects based on network proximity for Smart Cities. In general, the Physical Web is an approach for connecting any physical object to the web. With this approach, we can navigate and control physical objects in the world surrounding mobile devices. Alternatively, we can execute services on mobile devices, depending on the surrounding physical objects. Technically, there are different ways to enumerate physical objects. In this paper, we will target the models based on the wireless proximity.

*Keywords—proximity; Wi-Fi; Bluetooth; Smart Cities;*

## I. INTRODUCTION

The Physical Web is a term that describes the process of presenting everyday objects on Internet [1]. This approach offers mobile users the way to manage their daily tasks at using everyday objects. The (physical) objects become smart and remotely controllable. This model lets mobile users navigate and control physical objects in the world surrounding mobile devices. Also, it helps to perform everyday tasks depending on the surrounding physical objects. Of course, the key question is the way to enumerate physical objects.

The well-known IP address is an example of identification. Alternatively, Uniform Resource Identifiers (URIs), which include both locators and names, provide a higher-level concept for mapping those devices to existing Web technologies [2]. The classical model for the Physical Web links Uniform Resource Locators (URLs) with the objects. The Uniform Resource Locator (URL) is used in conjunction with a Distributed Name Service (DNS) to route requests and connect to services. A distinguishing aspect of the Physical Web (at least, as per Google's vision) is to consider URIs as the primary identifier.

We can mention in this context well known Quick Response (QR)-codes. QR-code is a 2D barcode, very often used for mapping URLs to physical objects [3]. Figure 1 presents the mapping for proximity measurements.

Wireless tags are one of the most often used approaches for physical objects markup. Wireless tags can support standard protocols like Bluetooth, Bluetooth Low energy (BLE) and Wi-Fi. The above-mentioned protocols are supported by almost all modern mobile phones.

So, for mobile devices (mobile users) the detection of tags is actually the detection of wireless nodes. Note, that in this approach other mobile devices can play a role of the tag too. And the network proximity (or wireless proximity) here describes data models based on the detection of proximity to the surrounding network nodes.

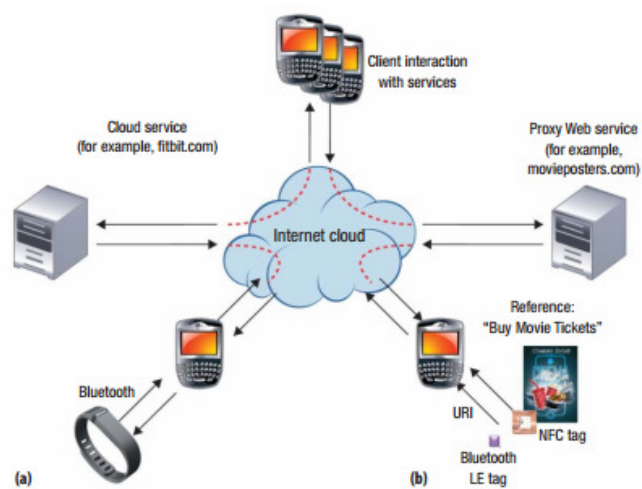

Fig. 1. Direct and proximity-based communications [1]

In this paper, we would like to discuss several approaches for building information systems (mobile services) based on the detection of physical objects via network proximity. These approaches are quite different from the traditional models of interaction with the physical objects. Usually, the interactions (the classical models for interactions) with physical objects are a subject of Internet of Things (Web of Things) [1,4]. But the interaction is not a mandatory requirement. Actually, we can present a lot of services, based on the presence of surrounding physical object (network tags in case of network proximity). The presented services do not incur two-way data exchange with the physical objects (with the network tags in case of network proximity). For all of them, it is enough to detect and identify the objects.

In general, the proximity is a very conventional way for context-aware programming in a mobile world [5]. There are many practical use cases, where the concept of the location can be replaced by that of proximity. Indoor services are one of the simplest examples. Proximity can be used as a main formation for context-aware browsers [6]. The context-aware browser lets mobile users discover (browse) pieces of data depending on the current context. And the context in our case is a set of visible (reachable) network nodes.

The usage of network proximity for context-aware programming (context-aware information systems) is very explainable. At his moment, network modules are most widely used "sensors" for mobile phones. All smartphones nowadays have Bluetooth, Bluetooth Low Energy, and Wi-Fi modules. So, Wi-Fi (Bluetooth) related measurements are included into standard interfaces of mobile operating systems. The above-mentioned measurements include the visibility for network nodes and signal strength. By the definition, the distribution area for Bluetooth signal, for example, is limited. So, as soon as some Bluetooth node is visible from a mobile device (a mobile phone, for example), then this device is somewhere nearby that node (it is so-called Bluetooth distance). The same is true for Wi-Fi access point. And this proximity information (network proximity) can replace location data. There are two main reasons for this replacement. At the first hand, we can target here all indoor application, where Global Positioning System (GPS) is not reliable.

For network proximity-based context-aware applications, any existing or even especially created wireless node could be used as a presence sensor that can play the role of a trigger. This trigger can open access to some content, discover existing content, as well as cluster nearby mobile users [7].

## II. ON PROXIMITY SERVICES

As the first use case, we can mention here a solution by SITA. It is a Community Service for Managing Global Airport iBeacon (Bluetooth Low Energy tags) Deployments [8].

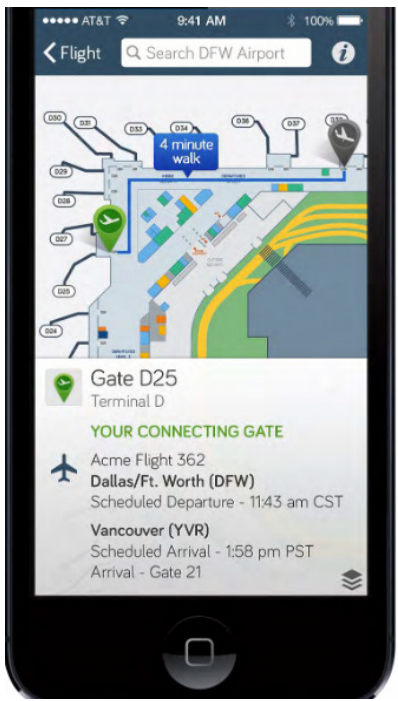

Fig. 2 Airport navigation application [9]

Figure 2 shows airport navigation and information mobile application based on the Beacon Registry. The Beacon Registry is a registry of common use iBeacons for the Air Transport Industry (ATI). It provides the following services:

It allows tags owners (airlines, airports or 3-rd parties) to manage their wireless tags infrastructure and track where they are placed in an airport.

It enables airports to monitor tags deployment to prevent radio interference with existing Wi-Fi access points

It provides beacons owners with a simple mechanism to set the 'meta-data' associated with beacons.

It provides an API for application developers who want to use these tags for developing travel and other related apps.

So, it is a common use database with wireless tags for airports worldwide. The common database lets developers provide information services for visitors nearby the tags.

Actually, indoor information (it is not necessary navigation) is a perfect example for proximity services. Obtaining GPS data indoor is not reliable and sometimes even impossible. In the same time, modern offices usually have plenty of wireless nodes. SITA uses dedicated wireless tags. But the same effect could be achieved with existing wireless networks nodes (Bluetooth nodes and Wi-Fi access points). But wireless nodes bring another interesting use case. The wireless node could be moveable. For example, we can create Bluetooth node right on the mobile phone. As soon as this phone is moved, the associated data "moved" too. Their visibility is linked to the current position of the phone. So, our context information will "follow" to the moved object. In this connection, we can mention CityProximus project (Ventspils University College, Moscow State University, Fraunhofer Fokus).

CityProximus includes the following elements:

- a crowd-sensing system for collecting a city-level database of wireless networks;

- a data persistence solution (database) for saving and maintaining wireless networks data;

- an editor for rule settings;

- a context-aware browser

For collecting wireless information data, this project uses a custom version of Funf package [10]. As per data persistence, visibility rules for context are organized with Apache Accumulo [11]. The simplest formation for proximity based rules is a visibility for the particular wireless node. In other words, each rule looks like this:

```
IF (Node_is_visible) THEN { activate some content }
```

Actually, iBeacons from Apple and Eddystone (dedicated Bluetooth tags) from Google follow this model. The visibility for any node could be described by its MAC-address. It means that any rule could be presented in the following form:

```
MAC-address -> content for activation
```

Rules will be described individually for the each wireless node. So, we have a key (MAC-address) and a vector of data chunks (texts, images, etc.) It is a typical key-value data model. This data model is one of the most used models for NoSQL approach. And Apache Accumulo is Open Source distributed key-value store.

Another interesting use case for context-aware services with wireless tags is associated with cars (and transport, in general). Many modern cars are wireless network nodes too. For example, many cars present Bluetooth node. Originally, it could be used for services procedures and access to multimedia centers. But without the connectivity (and without the associated security problems) we can treat such a node as a sensor. So, we can associate some data with our car. It could be some personal classified, commercial information, etc. Of course, such a "tag" is moveable. So, our linked data will follow to the tag (to the car, actually).

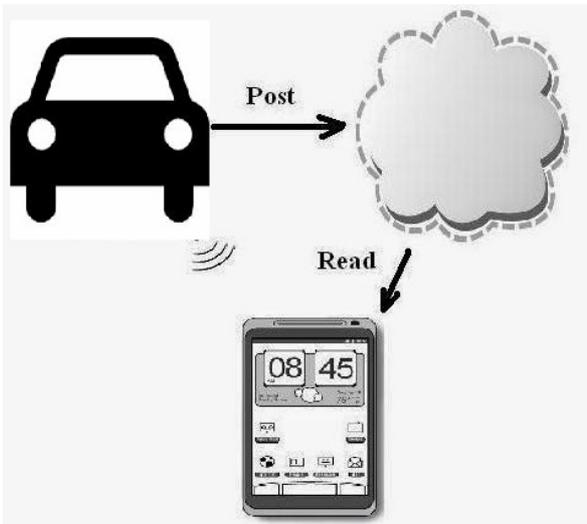

Fig. 3. Bluetooth Data Points [12]

In this case, mobile users (mobile devices, other cars) nearby the car with "linked" data will be able to obtain this information. Of course, it could be a simple notification for mobile readers, but it could start some actions in Machine to Machine (M2M) applications too.

III. ON MOBILE STATISTICS

As the next class of city-level services, associated with the proximity, we should mention the statistics. Tracking the presence of mobile users (subscribers) is one of the most interesting and useful sources of information in smart cities data processing [13]. Monitoring of mobile users (in fact, it is monitoring for mobile devices) supplies data to evaluate the mobility of residents, planning transport routes, etc. [14]. As the practical use cases, we can mention, for example, retail applications where analysis of the presence of mobile subscribers can be used to improve service, evaluation of marketing campaigns, planning, indoor navigation, data delivery in smart cities applications, etc. [15]. At this moment, we can list well known and commonly used methods for determining the location of mobile devices based on the location of Wi-Fi access points [4]. Mobile operating systems (mobile applications) can use the information about the objects of the network infrastructure for verifying (or even determine) the true state of the subscriber. By analyzing the signal strength and visibility of access points, we can build various metrics about the location of mobile devices (mobile users). For example, passive Wi-Fi monitoring is one of the common approaches. It lets anonymously collect data about mobile users (mobile devices) in a proximity of some access point. Collected data, in general, could be presented as some web-log. It is a direct analogue of web-log, collected by the web server. Collecting traces of Wi-Fi beacons is the well-known approach for getting the locations of mobile devices. Beacon frames are used to announce the presence of a Wi-Fi network. As a result, an 802.11 client receives the beacons sent from all nearby access points. The client receives beacons even when it is not connected to any network. In fact, even when a client is connected to some particular Access Point (AP), it periodically scans Wi-Fi channels to receive beacons from other nearby APs [16].

Data linked to wireless nodes bring another possibility to mobile statistics. The fact, that some particular tag (wires node) is scanned by mobile users is a direct analogue for web site request. Of course, it is anonymous, as an ordinary web request. Instead IP address, we have MAC-address. There is no Reference field (in web-log it is URL the request comes from), but there is a context-related information (a list of the visible wireless nodes). This information lets build some interesting services like "heat map" – usage (visiting) statistics for indoor. What is interesting also, this statistics could be used in real time. In other words, we add statistics to data visibility rules and incorporate these data right into context information. So, the output (visible information) will depend on the wireless reachable node and accumulated statistics.

IV. ON FUTURE DEVELOPMENT

In the nearest time, we will see a new wave of proximity based solutions. By our opinion, it will be associated with two - ways communication with tags (devices in tag's role). The closest example is Wi-Fi Aware program [17].

Wi-Fi Aware is a new Wi-Fi Alliance certification program that extends Wi-Fi's capabilities with a real-time and energy-efficient discovery mechanism. This mechanism provides continuous device-to-device discovery, even without a GPS, cellular or hotspot connection. Wi-Fi Aware lets applications continuously discover other devices and services within Wi-Fi range before making a connection. It turns proximity into a core function for Wi-Fi device. Wi-Fi Aware will make it easy to find information and services available in an area that match preferences set by the user. And what is especially important, it works well even in crowded environments. As per Wi-Fi alliance, Wi-Fi Aware will be a key enabler of a personalized social, local, and mobile experience, enabling users to find video gaming opponents, share media content, and access localized information all before establishing a connection. It allows bidirectional sharing of small pieces of data. So, the above-mentioned the physical web will be directly supported by devices.

We should note, that The Bluetooth Special Interest Group (SIG) officially announced the formation of the Bluetooth Smart Mesh Working Group. This working group will build the architecture for standardized mesh networking capability for Bluetooth Smart technology [18]. In general, mesh protocol allows Bluetooth to create a decentralized network of interconnected Bluetooth Smart devices. The proximity related tasks discussed above do not require the connectivity. But the process of forming mesh networks for Bluetooth devices could be very similar (conceptually) to Wi-Fi Aware. They both need some mechanism for continuous device-to-device discovery. So, the question is only how to allow developers access to this mechanism.